\documentclass[preprint]{emulateapj}

\slugcomment{Draft version}

\shorttitle{CANGAROO-III observation of HESS~J1614$-$518}
\shortauthors{Mizukami et al.}

\begin{document}


\title{CANGAROO-III observation of TeV gamma rays from the unidentified gamma-ray source HESS J1614$-$518} 


\author{
T.~Mizukami\altaffilmark{1},
H.~Kubo\altaffilmark{1},
T.~Yoshida\altaffilmark{2},
T.~Nakamori\altaffilmark{3},
R.~Enomoto\altaffilmark{4},
T.~Tanimori\altaffilmark{1},
M.~Akimoto\altaffilmark{5},
G.~V.~Bicknell\altaffilmark{6},
R.~W.~Clay\altaffilmark{7},
P.~G.~Edwards\altaffilmark{8},
S.~Gunji\altaffilmark{9},
S.~Hara\altaffilmark{10},
T.~Hara\altaffilmark{10},
S.~Hayashi\altaffilmark{11},
H.~Ishioka\altaffilmark{5},
S.~Kabuki\altaffilmark{1},
F.~Kajino\altaffilmark{11},
H.~Katagiri\altaffilmark{12},
A.~Kawachi\altaffilmark{5},
T.~Kifune\altaffilmark{4},
R.~Kiuchi\altaffilmark{13},
T.~Kunisawa\altaffilmark{4},
J.~Kushida\altaffilmark{5},
T.~Matoba\altaffilmark{2},
Y.~Matsubara\altaffilmark{14},
I.~Matsuzawa\altaffilmark{5},
Y.~Mizumura\altaffilmark{5},
Y.~Mizumoto\altaffilmark{15},
M.~Mori\altaffilmark{16},
H.~Muraishi\altaffilmark{17},
T.~Naito\altaffilmark{10},
K.~Nakayama\altaffilmark{4},
K.~Nishijima\altaffilmark{5},
M.~Ohishi\altaffilmark{4},
Y.~Otake\altaffilmark{9},
S.~Ryoki\altaffilmark{4},
K.~Saito\altaffilmark{5},
Y.~Sakamoto\altaffilmark{5},
V.~Stamatescu\altaffilmark{7},
T.~Suzuki\altaffilmark{2},
D.~L.~Swaby\altaffilmark{7},
G.~Thornton\altaffilmark{7},
F.~Tokanai\altaffilmark{9},
Y.~Toyota\altaffilmark{2},
K.~Tsuchiya\altaffilmark{18},
S.~Yanagita\altaffilmark{2},
Y.~Yokoe\altaffilmark{5},
T.~Yoshikoshi\altaffilmark{4},
Y.~Yukawa\altaffilmark{4}
}

\email{mizukami@cr.scphys.kyoto-u.ac.jp, kubo@cr.scphys.kyoto-u.ac.jp}

\altaffiltext{1}{ Department of Physics, Graduate School of Science, Kyoto University, Sakyo-ku, Kyoto 606-8502, Japan} 
\altaffiltext{2}{ Faculty of Science, Ibaraki University, Mito, Ibaraki 310-8512, Japan} 
\altaffiltext{3}{ Graduate School of Advanced Science and Engineering, Faculty of Science and Engineering, Waseda University, 3-4-1 Ohkubo, Shinjuku, Tokyo 169-0072, Japan} 
\altaffiltext{4}{ Institute for Cosmic Ray Research, University of Tokyo, Kashiwa, Chiba 277-8582, Japan} 
\altaffiltext{5}{ Department of Physics, Tokai University, Hiratsuka, Kanagawa 259-1292, Japan} 
\altaffiltext{6}{ Research School of Astronomy and Astrophysics, Australian National University, ACT 2611, Australia} 
\altaffiltext{7}{ School of Chemistry and Physics, University of Adelaide, SA 5005, Australia} 
\altaffiltext{8}{ CSIRO Astronomy and Space Science, Australia Telescope National Facility, Epping, NSW 2121, Australia} 
\altaffiltext{9}{ Department of Physics, Yamagata University, Yamagata, Yamagata 990-8560, Japan} 
\altaffiltext{10}{ Faculty of Management Information, Yamanashi Gakuin University, Kofu, Yamanashi 400-8575, Japan} 
\altaffiltext{11}{ Department of Physics, Konan University, Kobe, Hyogo 658-8501, Japan} 
\altaffiltext{12}{ Department of Physical Science, Hiroshima University, Higashi-Hiroshima, Hiroshima 739-8526, Japan} 
\altaffiltext{13}{Institute of Particle and Nuclear Studies, High Energy Accelerator Research Organization, 1-1 Oho, Tsukuba, Ibaraki, 305-0801, Japan} 
\altaffiltext{14}{ Solar-Terrestrial Environment Laboratory,  Nagoya University, Nagoya, Aichi 464-8602, Japan} 
\altaffiltext{15}{ National Astronomical Observatory of Japan, Mitaka, Tokyo 181-8588, Japan} 
\altaffiltext{16}{ Department of Physics, College of Science and Engineering, Ritsumeikan University, 1-1-1 Nojihigashi, Kusatsu, Shiga, 525-8577, Japan} 
\altaffiltext{17}{ School of Allied Health Sciences, Kitasato University, Sagamihara, Kanagawa 228-8555, Japan} 
\altaffiltext{18}{ National Research Institute of Police Science, Kashiwa, Chiba 277-0882, Japan}



\begin{abstract}
We report the detection, with the CANGAROO-III imaging atmospheric
Cherenkov telescope array, of a very high energy gamma-ray signal from
the unidentified gamma-ray source HESS~J1614$-$518, which was
discovered in the H.E.S.S.\ Galactic plane survey.  Diffuse gamma-ray
emission was detected above 760\,GeV at the 8.9~$\sigma$ level during
an effective exposure of 54\,hr from 2008 May to August. The spectrum
can be represented by a power-law:
$(8.2\pm2.2_{stat}\pm2.5_{sys})\times10^{-12}\times(E/1~$TeV$)^{-\gamma}$
cm$^{-2}$~s$^{-1}$~TeV$^{-1}$ with a photon index $\gamma$ of
$2.4\pm0.3_{stat}\pm0.2_{sys}$, which is compatible with that of the
H.E.S.S.\ observations.  By combining our result with multi-wavelength
data, we discuss the possible counterparts for HESS~J1614$-$518 and
consider radiation mechanisms based on hadronic and leptonic processes
for a supernova remnant, stellar winds from massive stars, and a
pulsar wind nebula. Although a leptonic origin from a pulsar wind 
nebula driven by an unknown pulsar remains possible, 
hadronic-origin emission from an unknown supernova remnant is preferred.
\end{abstract}




\section{Introduction}
Recent progress with Imaging Air Cherenkov Telescopes (IACTs)
is enabling the exploration of sites of cosmic-ray acceleration in our
galaxy. Very high energy (VHE) gamma rays are produced by the decay of
neutral pions which arise from interactions between the accelerated
protons and interstellar matter, or by Inverse Compton (IC) scattering
and Bremsstrahlung of high-energy electrons.  For example, VHE
gamma rays have been detected from young supernova remnants (SNRs)
such as RX J1713.7$-$3946 \citep{Muraishi_2000,can_1713,hess_1713}, RX
J0852.0$-$4622 \citep{can_0852,hess_0852,enomoto2}, and RCW86
\citep{Watanabe_2003,hess_rcw86}, which show possible evidence of
cosmic-ray acceleration 
\citep[e.g.,][]{can_1713,2005ApJ...624L..37M,Uchiyama_1713,tanaka_1713,yamazaki_2009}.
In addition, detections of VHE gamma rays from pulsar wind nebulae
(PWNe) such as the Crab nebula \citep{Weekes} and Vela X nebula
\citep{velax,enomoto} have shown that PWNe also play an important role
in particle acceleration in the Galaxy.  Recently, VHE gamma-ray
emission related to massive stars such as the X-ray binary systems
PSR~B1259$-$63 \citep{1259} and LS~I~+61 303 \citep{LSI}, 
the young open stellar clusters Cyg~OB2
\citep{2032}, Westerlund~2 \citep{wes2}, and Westerlund~1 \citep{2010ASPC..422..265O} 
have also been reported.
Moreover, the Galactic plane survey performed by the
H.E.S.S.\ observatory \citep{hess,hess2} discovered seventeen
unidentified VHE gamma-ray sources, including HESS~J1614$-$518.
Today, unidentified sources are the largest class of the 123
discovered VHE gamma-ray sources, most of which are located in the
Galactic plane (e.g., \citet{2005A&A...431..197A,milagro,hess_unid}; the TeVCAT catalog, \citet{TeV_catalog}, is a useful 
up-to-date, on-line resource).  In general, the lack of
non-thermal electromagnetic radiation from the radio to the X-ray
bands may be evidence of hadron acceleration because the IC scenario
requires a lower magnetic field than the typical interstellar magnetic
field intensity of a few $\mu$G.  Revealing the possible radiation
mechanism(s) of each unidentified source is therefore important for
identifying the origin(s) of cosmic rays.

H.E.S.S.\ reported that HESS~J1614$-$518 had a high flux level, 25\%
of the Crab nebula, above 200\,GeV with a photon index of 2.4 and an
elliptical morphology with a semi-major axis of $14\pm1$~arcmin and a
semi-minor axis of $9\pm1$~arcmin \citep{hess2}. The peak position has
an offset of 8.7~arcmin to the north-east from the central position.
\citet{landi2} and \citet{rowell} pointed out that HESS
J1614$-$518 may be associated with the 40~Myr-old young open star
cluster Pismis~22 \citep{piatti_pismis} which is located within the
VHE gamma-ray emission region at a distance of 1.0$\pm$0.4\,kpc
and has sufficient luminosity to produce the observed gamma-ray
luminosity, assuming 20\% energy conversion from the stellar winds of ten
B-type stars.  However, there are several issues in identifying
HESS~J1614$-$518 with Pismis~22 since the size of Pismis~22,
2.0\,arcmin in diameter, is one order of magnitude smaller than the
VHE gamma-ray emission size and the location has a 12\,arcmin offset
from the VHE gamma-ray emission peak.  In addition, there has been no
detailed discussion of the radiation mechanism.

The X-ray satellite {\it Suzaku} observed this region with the X-ray
imaging spectrometer (XIS) and found three X-ray sources in their
follow-up observation in 2006 \citep{matsumoto}.  One of these, called
Suzaku source~A, is located very close to the VHE gamma-ray peak
position with an offset of 0.8\,arcmin. 
The spectrum is well fitted by a single power-law model with a photon
index of $1.73^{+0.33}_{-0.30}$ and a hydrogen equivalent column
density of $1.21^{+0.50}_{-0.41}\times10^{22}$ cm$^{-2}$.  The
distance to Suzaku source~A is approximately 10~kpc, which was derived
from the hydrogen equivalent column density using the total Galactic
H\,{\sc i} column density toward HESS~J1614$-$518 of
$\sim2.2\times10^{22}$~cm$^{-2}$ \citep{Dickey}.  The size of the
X-ray emission region is slightly larger than the {\it Suzaku} Point
Spread Function (PSF) of 1.8\,arcmin and smaller than the size of the
VHE gamma-ray region.  This difference is seen in PWNe such as
HESS~J1825$-$137 \citep{hess3,uchiyama_1825} and Vela X
\citep{velax_X,velax} and could be explained by the difference between
the synchrotron cooling time of the electrons that radiate X-rays and those
that produce TeV gamma rays.
Electrons with an energy of 100\,TeV radiate X-rays and immediately
lose their energy by synchrotron cooling (e.g., the energy-loss
timescale is $\sim10^{2}$~yr assuming a magnetic field of 20~$\mu$G
\citep{Sturner_1997}), while electrons with an energy of 1\,TeV, which
are responsible for the VHE gamma ray emission through IC scattering, are more
slowly cooled by synchrotron radiation (e.g., the energy-loss
timescale is $\sim10^{4}$~yr assuming the same parameters as above)
and can travel further from their source.  However, the ratio between
the observed VHE gamma-ray and X-ray fluxes, F(1--10 TeV) / F(2--10
keV) of $\sim$34, is much larger than those of known PWNe ---
$2.6\times10^{-3}$, 0.7, and 1.5 for the Crab, MSH~15-52, and Vela X,
respectively \citep{Gaensler_1999, x_crab,Gaensler_2002, Dodson_2003,
  hess_crab,Aharonian_2005, velax, Manzali_2007, Nakamori_2008}.
Nevertheless, recent studies of HESS~J1640$-$465 \citep{funk} and
HESS~J1804$-$216 \citep{Higashi_2008} claim that this large ratio can
be explained by a time-evolving electron injection model, in which the
number of electrons injected into space by the pulsar decreases
proportionally to the spin-down of the pulsar.  On the other hand,
this large ratio is also expected in an old SNR with an age of
$\sim10^{5}$~yr, because of the difference between the cooling times
of electrons and protons \citep{yamazaki_old}.  We therefore discuss both a
PWN scenario and an SNR scenario in this paper.

Suzaku source~B is positioned towards the center of HESS~J1614$-$518
and is coincident with the position of Pismis~22. Since the hydrogen equivalent column density derived from the Suzaku spectrum 
is $(1.1\pm0.21)\times10^{22}$ cm$^{-2}$, which is comparable with that
of Suzaku source~A, Suzaku source~B may lie at a similar distance as for
Suzaku source A. This source has a non-thermal X-ray emission with a photon index of $3.19\pm0.32$. 
This soft index and X-ray
luminosities of $7.7\times10^{34}$~ergs~s$^{-1}$ and
$4.5\times10^{35}$~ergs~s$^{-1}$ in the 2$-$10~keV and 0.5$-$10~keV
ranges, respectively, assuming a distance of 10~kpc, are typical values
for an anomalous X-ray pulsar (AXP) \citep{AXP_01,AXP_02}. The
possible existence of the AXP also suggests this source as an SNR,
since AXPs are usually associated with SNRs, e.g., 1E~2259+586 with
CTB~109 \citep{AXP_01} and 1E~1841$-$045 with Kes~73
\citep{AXP_1841}. Since the position of this source is coincident with
Pismis~22, there is a possibility that this emission originates from
the stellar winds from the stellar cluster. Non-thermal X-ray emission
from a stellar cluster was reported from Westerlund~1
\citep{Muno_2006}, and TeV gamma rays were recently detected from this object
\citep{Ohm_2009}. However, this positional correlation may be only a
chance coincidence since the estimated distances to Suzaku source~B
and Pismis~22 are different by an order of magnitude. 
Although Suzaku source~B might be marginally extended, 
it is difficult to quantitatively estimate the spatial 
extension with the Suzaku PSF of 1.8~arcmin. 
If Suzaku source B is actually extended, additional scenarios besides an SNR
could be considered, e.g., a PWN from a pulsar/AXP as discussed in 
\citet{matsumoto}, or emission from the unresolved hot stars in Pismis 22.

The other source, Suzaku source~C, is a late B-type star as described in 
\citet{matsumoto}, and thus is
not a possible counterpart of HESS J1614$-$518.  

{\it Swift} observed this region with the X-ray telescope (XRT)
and found six X-ray sources (hereafter Swift sources~1 to 6)
\citep{landi1,landi2}.  All these sources were point-like and no
diffuse emission was found.  Two sources, Swift sources~1 and 4, are
located within the field of view (FOV) of the {\it Suzaku}
observation.  Swift source~1 is located close to Pismis~22 with an
offset of 42~arcsec.  This source is also coincident with Suzaku
source~B. Swift source~4 is coincident with Suzaku source~C.  Swift
sources~1, 2, 3, 5 are probably stars, while the nature of Swift
sources~4 and 6 were not identified, probably due to the poor
statistics.  Although Suzaku source~A was located in the FOV of the
{\it Swift} XRT, it was not detected with {\it Swift} probably due to
the limited exposure time ($\sim$1700~s) and/or the small effective
area.

The Fermi-LAT collaboration \citep{fermi_cat} reported the detection
of gamma rays in the 100\,MeV to 100\,GeV band from 1FGL~J1614.7-5138c
positioned 2.7\,arcmin away from the peak position of the VHE
gamma-ray emission.  In the radio band, no counterpart has been found
in the HESS~J1614$-$518 region; there is no enhancement in the 843~MHz
band, where the rms noise level is $\sim$2\,mJy\,arcmin$^{-2}$
\citep{radio,Murphy_2007}.  In this paper, we present TeV gamma-ray
observations of HESS~J1614$-$518 with the CANGAROO-III telescopes and
discuss the possible counterpart and the radiation mechanism by
considering multi-wavelength observations.

\section{Observations}

CANGAROO-III is an array of four IACTs (T1, T2, T3, and T4), located
at Woomera, South Australia (136$^{\circ}$47E, 31$^{\circ}$06S, 160 m
a.s.l.) \citep{Enomoto_2002b}. The oldest telescope, T1, which was the
CANGAROO-II telescope, has not been in use since 2004 due to its
smaller FOV and higher energy threshold. Each telescope has a 10~m
diameter reflector which consists of 114 segmented fiber reinforced
plastic spherical mirrors mounted on a parabolic frame
\citep{can_mir}. The imaging camera system consists of 427
photomultipliers (PMTs) and has a FOV of 4.0~deg \citep{can_cam}. The
PMT signals are digitized by charge analog-to-digital converters
(ADCs) and multi-hit time-to-digital converters (TDCs)
\citep{can_daq}. The observations were carried out from 2008 May to
August using a wobble mode in which the pointing position
was shifted both in declination and right ascension between
$\pm0.5$\,deg from the target position every 20 minutes
\citep{Nakamori_2008}. The target position was (R.A.,
decl. [J2000])=(243$^{\circ}$.579,$-$51$^{\circ}$.820) which is the
center of the source position reported by H.E.S.S.

The light-collecting efficiency, including the reflectivity of the
mirror segments and the light guides, and the quantum efficiency of PMTs,
was monitored by a muon-ring analysis with individual trigger data
taken in the same period \citep{enomoto}. The average quantity of
light per unit arc-length of muon rings is approximately proportional
to the light-collecting efficiency. We also did not use the second
oldest telescope, T2, since it was very difficult to calibrate with
muon-ring data to calculate the efficiency at this time due to the
deterioration of mirror reflectivity \citep{Enomoto_2009}. The two
telescopes T3 and T4 were used. From the muon-ring analysis for the
data taken in this period, the light-collecting efficiency of each
telescope, which is used in the Monte Carlo simulations, with respect
to the original mirror production time, was 0.58 and 0.50 for T3 and
T4, respectively. We reject the data for which the average
trigger-rate over a one minute period was under 5\,Hz to remove data
taken in cloudy conditions. The effective exposure time amounts to
53.6\,hr and the energy threshold was 760~GeV.

\section{Analysis}

The standard analysis of the CANGAROO-III collaboration
\citep{enomoto,kabuki} was applied to the data. The calibrations for
the cameras and ADCs were carried out daily using LEDs. After calibration,
the recorded charges of each pixel in the camera were converted to the
number of photo-electrons. Time-walk corrections for TDC data were
carried out using data taken by changing the luminosity of the
LEDs. After calibration, every shower image was cleaned through the
following criteria. Only pixels that had $\ge$5.0 photo-electrons were
used as ``hit pixels". Clusters of five or more adjacent hit pixels
with arrival times within 30\,nsec of the average hit time of all
pixels were recognized as a shower cluster. Before calculating image
moments --- the ``Hillas parameters" \citep{hillas} --- we applied an
``edge cut" to the data \citep{enomoto2} and rejected events having
hit pixels in the outer-most layer of the camera. The orientation
angles were determined by minimizing the sum of the squared widths
with a constraint given by the distance predicted by Monte Carlo
simulations.

In order to derive the gamma-ray likeliness, we used the Fisher
Discriminant method \citep{fisher,enomoto}. The input parameters were
\begin{center}
$\vec{P}=(W_{3},W_{4},L_{3},L_{4})$,
\end{center}
where $W_{3},W_{4},L_{3},L_{4}$ are energy-corrected $Widths$ and
$Lengths$ for T3 and T4 camera images. The Fisher Discriminant
(hereafter FD) is defined as $FD=\vec{\alpha}\cdot\vec{P}$, where
$\vec{\alpha}$ is a set of coefficients mathematically determined in
order to maximize the separation between two FD distributions for
gamma rays and hadrons.

\begin{figure}[!t]
\centering
\includegraphics[width=\linewidth]{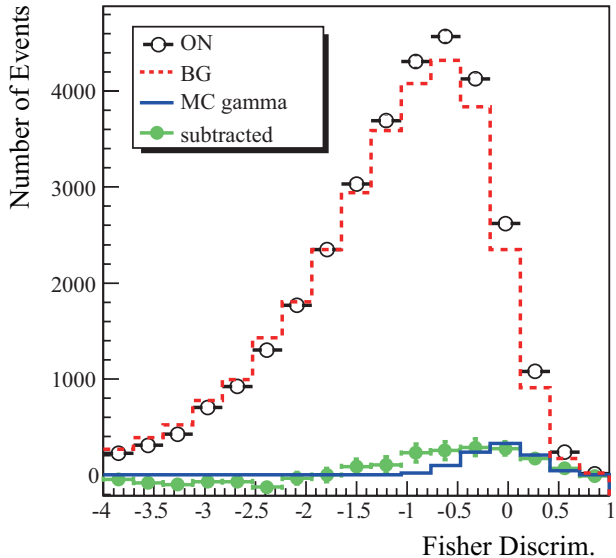}
\caption{FD distribution. The black circles show the FD obtained from the ON source region, $\theta^{2}\le0.2$~deg$^{2}$. The red and blue lines are the background and gamma-ray component estimated by the fit procedure described in the text. The green circles are obtained by the subtraction of the background from the ON source region.}
\label{fig_fd}
\end{figure}

To calculate the background, we selected a ring region around the
target, $0.3\le\theta^{2}\le0.5$ deg$^{2}$, where $\theta$ is the
angular distance to the center of HESS~J1614$-$518
\citep{Higashi_2008}. We obtained the FD distributions of hadrons
$F_{b}$ from this region and gamma rays $F_{g}$ from Monte Carlo
simulations. Finally, we were able to fit the FD distribution of the
events from the target with a linear combination of these two
components. The observed FD distributions $F$ were represented by
\begin{center}
$F=\beta F_{g} + (1-\beta)F_{b}$,
\end{center}
where $\beta$ is the ratio of gamma-ray events to the total number of
events. Here $\beta$ is a fitting parameter and the obtained FD
distributions are shown in Figure~\ref{fig_fd}.

\section{Results}
\begin{figure}[!t]
  \centering
  \includegraphics[width=\linewidth]{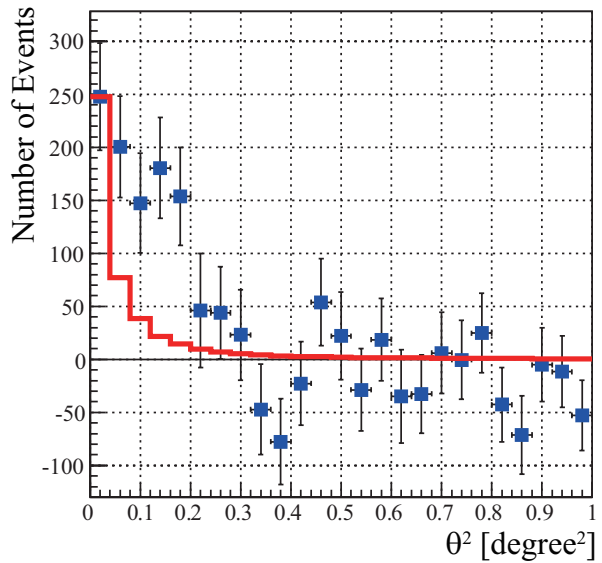}
  \caption{The $\theta^{2}$ plot, where $\theta^{2}=0$ corresponds to the fitted center of gravity of HESS~J1614$-$518 from H.E.S.S.\ \citep{hess2}. The blue data points represents the excess events in each $\theta^{2}$ bin and the red solid line represents our PSF derived from the Monte-Carlo simulation.}
  \label{fig_q2}
 \end{figure}

\begin{figure}[!t]
\centering
\includegraphics[width=\linewidth]{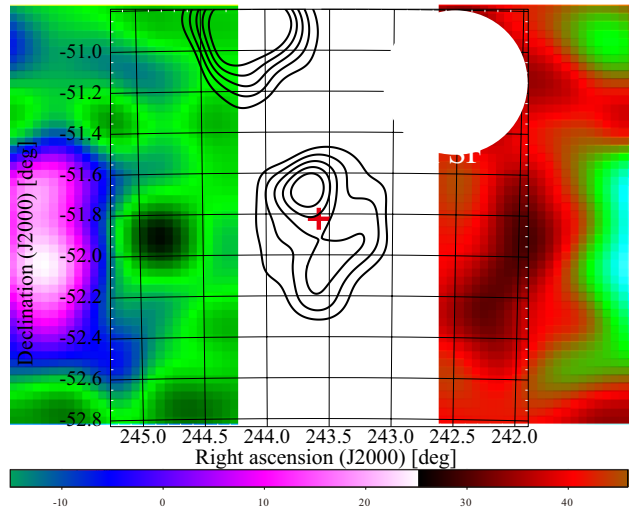}
\caption{Morphology of gamma-ray--like
events. The number of excess events per $0^{\circ}.04 \times 0^{\circ}.04$ cell is
smoothed by a Gaussian with $\sigma=0.24$~degree, which is the CANGAROO-III PSF, and plotted in equatorial coordinates. The black solid contours show the VHE gamma-ray emission seen by H.E.S.S. 
Lines correspond to 20, 30, 40, 50, \& 60 gamma-ray counts. 
The red cross shows the H.E.S.S.\ center of gravity of HESS~J1614$-$518 \citep{hess2}.}
\label{fig:map}
\end{figure}

\begin{figure}[!t]
  \centering
  \includegraphics[width=\linewidth]{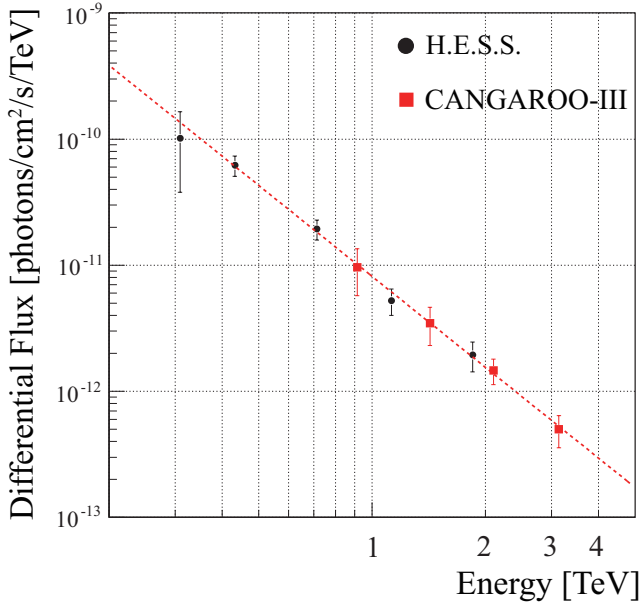}
  \caption{Differential flux of HESS~J1614$-$518. Squares and circles show the CANGAROO-III and the H.E.S.S.\ data points, respectively. The best fit power-law from this work is shown by the dotted line.}
  \label{fig_df}
 \end{figure}

%
%
The obtained $\theta^{2}$ plot is shown in Figure~\ref{fig_q2} with the PSF of our telescopes. Above 760\,GeV we detected 950$\pm$107 excess events within $\theta^{2}\le0.2$~deg$^{2}$. 
%
%
The morphology of gamma-ray--like events, obtained from a gaussian
smoothing with the CANGAROO-III PSF of 0.24\,deg, is shown in
Figure~\ref{fig:map}. The extent of the VHE gamma-ray emission was
estimated by a 2D Gaussian fit on our excess map. The obtained
standard deviation was $0.44\pm0.03$~deg which is broader than the
CANGAROO-III PSF. The centroid position was determined to be (R.A.,
decl.~[J2000])$=(243^{\circ}.634$, $-51^{\circ}.950)$. The offset from
the best-fit position reported by H.E.S.S.\ is ($\triangle$R.A.,
$\triangle$decl.)=($0^{\circ}.055\pm0^{\circ}.018$,$-0^{\circ}.130\pm0^{\circ}.033$).
The offset is not significant compared with our PSF. A systematic
difference due to the difference in energy thresholds between
H.E.S.S.\ and CANGAROO-III may also contribute to this offset.
%
%
The reconstructed VHE gamma-ray differential flux is shown in
Figure~\ref{fig_df} together with the H.E.S.S.\ measurements.  The
spectrum is compatible with a single power-law:
$(8.2\pm2.2_{stat}\pm2.5_{sys})\times10^{-12}\times(E/1~$TeV$)^{-\gamma}$~cm$^{-2}$~s$^{-1}$~TeV$^{-1}$
with a photon index $\gamma$ of $2.4\pm0.3_{stat}\pm0.2_{sys}$.  The
relevant systematic errors are due to the atmospheric transparency,
night-sky background fluctuations, uniformity of camera pixels, and
light-collecting efficiencies. In addition, to estimate the systematic
error due to the size of signal integration region, we changed the
region from $\theta ^{2} < 0.14$~deg$^{2}$ to 0.30~deg$^{2}$, which
was included in the systematic error.  For comparison with the
ring-region background, we took background events from opposite
positions of HESS J1614$-$518 observations in the wobble mode, and then obtained a
differential flux of
$(6.4\pm2.0_{stat}\pm2.4_{sys})\times10^{-12}\times(E/1~$TeV$)^{-\gamma}$~cm$^{-2}$~s$^{-1}$~TeV$^{-1}$
with a photon index $\gamma$ of $2.4\pm0.6_{stat}\pm0.3_{sys}$, which
was consistent with that derived with the ring-region background.
The VHE gamma-ray extension, centroid position, and flux obtained with CANGAROO-III are consistent with results from H.E.S.S. 
This result suggests that the VHE gamma-ray emission was unchanged between 2004 and 2008.

\section{Discussion}

\begin{figure}[!t]
\centering
\includegraphics[width=\linewidth]{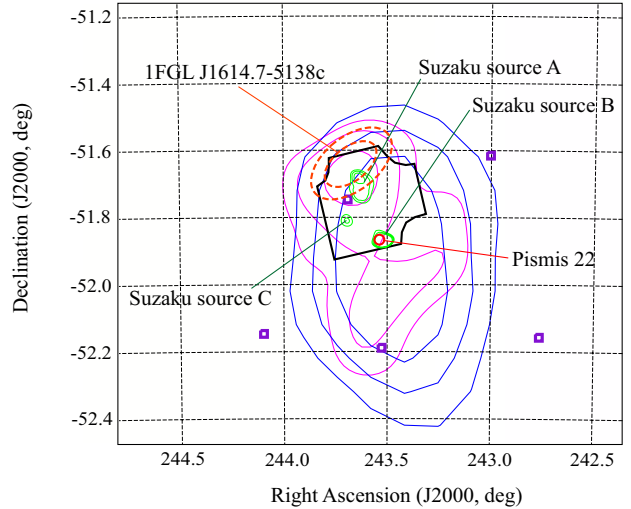}
\caption{Morphological relationship between the X-ray and gamma-ray observations. Blue, magenta, and green contours show the emission regions of CANGAROO-III, H.E.S.S., and {\it Suzaku}, respectively. Orange dashed circles show the 68\% and 90\% error ellipses of the position of the {\it Fermi} source. The red circle shows the Pismis~22 position and purple squares show the positions of nearby pulsars \citep{Manchester,Hobbs_2004}. The bold black line shows the observed region by the {\it Suzaku} XIS in \citet{matsumoto} excluding the calibration source region on the corners.}
\label{fig:map3}
\end{figure}

We now discuss the plausible radiation mechanisms of HESS~J1614$-$518 using
the results of CANGAROO-III, H.E.S.S., {\it Fermi}, and {\it Suzaku}
observations. Since the spectra of {\it Swift} sources were not
available, we did not use the {\it Swift} data. Figure~\ref{fig:map3}
shows the morphological relationship between each observation. The
non-thermal X-ray emission from Suzaku source~A is positioned very
close to the H.E.S.S.\ gamma-ray peak, the position of the {\it Fermi}
source 1FGL J1614.7-5138c, and within the emission region detected
with CANGAROO-III. Thus, this could be the most likely counterpart for
HESS~J1614$-$518.  We note here that since the FOV of the {\it Suzaku}
observation covered only the part of the TeV gamma-ray emission region, 
as shown in Figure~\ref{fig:map3}, the current observed X-ray flux may
only be a fraction of the entire X-ray emission from the entire region
of the VHE gamma-ray emission. To discuss the emission mechanism with
more accuracy, further X-ray observations of the entire region of VHE
gamma-ray emission are needed. For further constraints on emission 
models, we tried to derive the flux at 8~$\mu$m at the
Suzaku source A position, from the archival data with the Infrared 
Array Camera (IRAC; \citet{spitzer}) onboard the {\it Spitzer} Space 
Telescope. However, since there was contamination from a nearby source, 
we obtained only an upper limit. 

\noindent
{\bf SNR Scenario:} Suzaku source~B, which may be related to a possible AXP,
is positioned roughly in the center of the VHE gamma-ray emission.  We
thus postulate a scenario in which a supernova explosion occurred at
the position of Suzaku source~B and the shock of the SNR has now
reached the position of Suzaku source~A, emitting both the X-ray and
gamma-ray emission.

\noindent
{\bf PWN Scenario:} A PWN could also emit diffuse gamma-ray
emission. Five pulsars have been found in this region,
PSR~J1611$-$5209, PSR~J1612$-$5136, PSR~J1613$-$5211,
PSR~J1614$-$5144, and PSR~J1616$-$5208
\citep{Manchester,Hobbs_2004}. As described in section~1, 
Suzaku source B would be a PWN 
if associated with a pulsar or AXP.
The smaller size of Suzaku source~A or source~B
compared to that of the VHE gamma-rays also appears in other PWNs
because of synchrotron cooling (section~1). 

We will discuss the PWN scenario and the associated pulsar which could supply
enough particles to reproduce both the X-ray emission of Suzaku
source~A or source~B and the VHE gamma-ray emission.

\noindent
{\bf Stellar Wind Scenario:} The young open cluster Pismis~22 is
located towards the center of HESS~J1614$-$518 and is also a possible
counterpart. Its age is $\sim4.0\times10^{7}$\,yr and the distance is
1$\pm$0.4\,kpc from the Earth. The coincidence between a young open
cluster and a VHE gamma-ray source is also seen in Westerlund~2 and
Cyg OB2 as described in section~1. Stellar winds from massive stars
could form a shock front, accelerate charged particles, and produce
the high energy radiation \citep{Voelk_1982,Bednarek}.

Another possibility is an association between an SNR, PWN and the open
cluster, since Pismis~22 is old enough for some massive stars to
finish their life as supernovae. In addition, as described in
section~1, binary systems that emit VHE gamma rays also have been
discovered. However, HESS~J1614$-$518 does not seem to be associated with
a binary system, since all sources of this type have point-like
emission, which is in contrast to the results of CANGAROO-III and
H.E.S.S. We discuss the above scenarios in detail in the following
subsections.

\subsection{SNR Scenario}

In the SNR scenario, we assume that the X-rays from Suzaku source~A
and the VHE gamma rays are emitted by charged particles accelerated by
the shock in the SNR shell, and Suzaku source~B is an associated AXP
which is positioned in the center of the SNR shell. Additionally, we
will discuss the possible correlation between the SNR and Pismis~22.

\begin{figure}[!t]
  \centering
  \includegraphics[width=\linewidth]{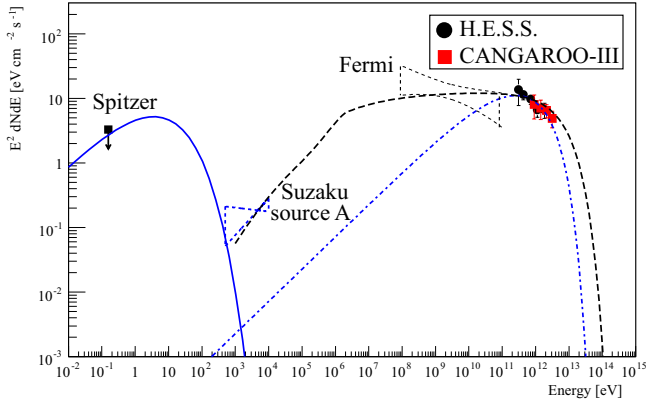}
  \caption{SED with leptonic model curves for HESS J1614$-$518. The dash-dotted and solid blue lines show IC and synchrotron emission derived from the single power-law electron spectrum with an exponential cutoff to fit the VHE emission with a magnetic field of 6~$\mu$G. The dashed black line shows a Bremsstrahlung curve for a number density of ambient matter of $600~p\rm{~cm}^{-3}$.}
  \label{fig:single_lep}
 \end{figure}

%
%
First we examine a leptonic model to explain the observed SED
(Fig. \ref{fig:single_lep}). For the X-ray spectrum, 
we use the {\it Suzaku} spectrum of Suzaku source~A 
correlated with the VHE gamma-ray peak, with a statistical error 
at the 90\% confidence level. For the {\it Fermi} spectrum, 
we used a 2$\sigma$ statistical error
and systematic errors of 1.8$\sigma$ in the flux and 1.2$\sigma$ in
the photon index \citep{fermi_cat}. 
Figure~\ref{fig:single_lep} shows the simple one-zone
leptonic model curves for HESS~J1614$-$518. Here we calculated
synchrotron, inverse Compton, and Bremsstrahlung model curves for a
single power-law with an exponential cutoff electron spectrum,
$dN_{e}/dE_{e}=K_{e}E_{e}^{-\Gamma_{e}}\exp(-E_{e}/E_{max\_e})$, where
$K_{e}$ is the normalization factor, $E_{e}$ is the electron energy,
$\Gamma_{e}$ is the spectral index of the electrons, and $E_{max\_e}$
is the maximum accelerated electron energy. 
To calculate the IC radiation, we used cosmic-ray microwave background and an interstellar
radiation field (ISRF) derived from the GALPROP package (v50p)
\citep{galprop_1,galprop_2} to estimate the seed photon field around
the HESS~J1614$-$518 region. Energy densities of 1.4\,eV~cm$^{-3}$ and
5.5\,eV~cm$^{-3}$ for IR and optical light were obtained,
respectively. In addition, this radiation field changed by less than
an order of magnitude when we varied the distance from 1\,kpc to
10\,kpc, with values in the range from 0.9 to 1.7~eV~cm$^{-3}$ and 1.1
to 5.5 eV~cm$^{-3}$ for IR and and optical light, respectively. We
fixed the power-law index to $\Gamma_{e}=2.0$ and fitted the VHE
gamma-ray spectrum by the IC emission. The maximum energy and total
energy of the electrons obtained were 4.2$\pm$1.5~TeV and $1.9 \times
10^{49} \times (d/ 10 {\rm kpc})^{2}$\,ergs. From comparison between 
the synchrotron model and the {\it Spitzer} upper limit, 
an upper limit of a magnetic field is determined to be 6~$\mu$G. 
%
%
The harder and fainter spectrum in the X-ray band compared to the VHE
gamma-ray spectrum did not allow the observed X-ray and VHE gamma-ray
spectra to be produced with synchrotron and IC emission, respectively,
by a single power law distribution of electrons.
There is also the possibility that Bremsstrahlung produces both the
X-ray and VHE gamma-ray emission \citep{Uchiyama_brems}, as shown in
Figure~\ref{fig:single_lep}.
This model gives a good reproduction for an ambient matter density
$n_{p}$ of 600~{\it p}~cm$^{-3}$. However, \citet{rowell} reported
that they found no obvious overlapping molecular clouds across a range
of inferred distances up to $\sim$6~kpc, with the NANTEN
$^{12}$CO(J=1-0) survey data \citep{Matsunaga}. 
In addition, we estimated an ambient matter density from the 
velocity-integrated data of the CO survey \citep{dame} 
to be $\sim$80 or 8~{\it p}~cm$^{-3}$ for 10 kpc and 1 kpc, 
respectively. Thus, the Bremsstrahlung model 
that requires an ambient matter density of 600~{\it p}~cm$^{-3}$ was
rejected.
Furthermore, there is difficulty in explaining the morphological 
difference between the
X-ray and the VHE gamma-ray emission because it requires an unlikely
situation in which the relatively high energy (multi-TeV) electrons
that are responsible for the VHE gamma-ray emission are distributed
over a more extended region than the relatively low energy (multi-keV)
electrons which are responsible for the X-ray emission.

We also checked that our estimate of the non-thermal synchrotron flux did not violate the thermal optical emission. The intrinsic optical background flux 
over the solid angle for the gamma-ray emission region 
$\sim 5\times 10^5$ eV cm$^{-2}$s$^{-1}$ is obtained from 
$U_{opt} (c/4 \pi) \Omega$, where $U_{opt}$ is the energy density 
of 5.5 eV cm$^{-3}$ for the optical light, $c$ is the speed of the
light, and  $\Omega \sim \pi(0.2 \times \pi/180)^2$ sr is the solid angle 
for the region. The optical background radiation is 10$^5$ times 
larger than the optical synchrotron flux for a magnetic 
field of 6~$\mu$G (Fig.~\ref{fig:single_lep}). 

\begin{figure}[!t]
  \centering
  \includegraphics[width=\linewidth]{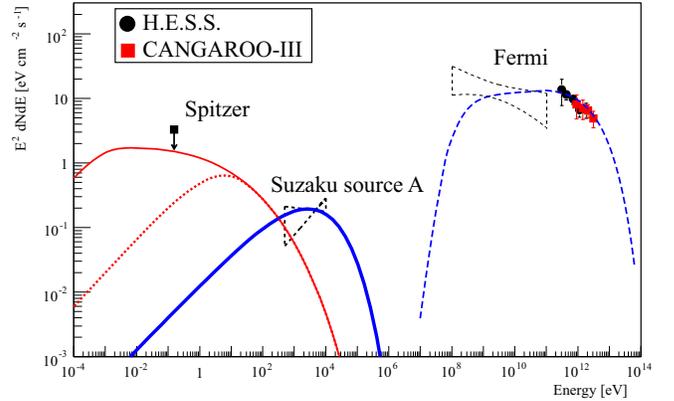}
  \caption{SED and the model curve for neutral pion decay (blue dashed line). The blue bold-solid line shows the synchrotron model curve for the primary electrons. The red solid and dotted lines show synchrotron model curves for the secondary electrons for a distance of 10~kpc and 1~kpc, respectively.}
  \label{fig:hadron}
 \end{figure}

%
%
Second, we examined a neutral-pion decay model. Based on a model
proposed by \citet{yamazaki_old}, only nucleonic particles remain in
an old SNR with an age of $\sim10^{5}$~yr, while primary electrons
have already lost most of their energy by the synchrotron cooling.
Figure~\ref{fig:hadron} shows the SED with the assumption that 
the population of accelerated protons can be expressed by a single 
power-law with an exponential cutoff, $dN_{p}/dE_{p} =K_{p} 
E_{p}^{-\Gamma_{p}}\exp(-E_{p}/E_{max\_p})$. We set the power-law 
index to $\Gamma_{p} = 2.0$.
%
%
The best-fit cutoff energy was obtained to be $E_{max\_p}=36 \pm
18$~TeV.  The total energy of high energy protons was calculated to be
$W_{p}
=1.2\times10^{52}(n_{p}/1~p\rm{~cm}^{-3})^{-1}(d/10\rm{kpc})^{2}$~ergs.
By setting $n_{p}=100~p$~cm$^{-3}$, the efficiency of energy
conversion to accelerate protons is 10\% for a typical total supernova
explosion kinetic energy of $\sim10^{51}$\,ergs. As described above,
no obvious molecular cloud was found in the NANTEN $^{12}$CO(J=1-0)
survey data. 
In addition, an ambient matter density 
from the velocity-integrated data of the CO survey 
(Dame et al., 2000) was estimated to be $\sim$80 or 8~{\it p}~cm$^{-3}$
for 10 kpc and 1 kpc, respectively. 
Thus, an assumption of 100~{\it p}~cm$^{-3}$ is likely 
for 10 kpc, but not for 1 kpc. 
Further observations 
are necessary to investigate the validity for such assumed density. 
%
%
Assuming the spectral index, $\Gamma_{p}=\Gamma_{e}=2.0$, the maximum
energy of primary electrons is determined to be $E_{max\_e}\le
E_{max\_p}$. The turnover energy of synchrotron emission is determined
from $E_{turn} =22$~keV~$\times (E_{max\_e}/50~{\rm TeV})^{2}
\times (B/200 \mu {\rm G})$. Since the hard index of the X-ray
spectrum required $E_{turn} \ge 10$~keV, a lower limit for the magnetic
field of $B\ge $200~$\mu$G was determined from this equation. The
model curve for this lower limit condition was shown in
Figure~\ref{fig:hadron}. From this magnetic field value, we set a
lower limit to the number ratio of protons to primary electrons,
$K_{pe}=K_{p}/K_{e}\ge 2.1\times 10^{5}
(n_{p}/1~p\rm{~cm}^{-3})^{-1}$.

%
%
We also calculated the contribution of emissions from secondary
electrons from {\it p-p} interactions between the same proton
population as above and the ambient matter density of
100\,$p$\,cm$^{-3}$. We followed the calculation in
\citet{Kelner_2008} to derive the spectrum of the secondary
electrons. 
Assuming a distance of 10\,kpc from the Earth, 
the distance between Suzaku source~A and source~B was calculated to be 35~pc.
Thus, we assumed that the radius of the SNR is 35~pc 
and roughly estimated the age of the SNR using the
equation~(2) in \citet{yamazaki_old} to be $3\times 10^{4}$~yr. 
Thus, the emissions from the secondary electrons were derived by
assuming continuous injection of electrons produced by a constant
proton spectrum over $3\times 10^{4}$~yr with a magnetic field of
200~$\mu$G \citep{Atoyan_1999}. The obtained synchrotron curve is also
shown in Figure~\ref{fig:hadron}. The obtained inverse Compton and
Bremsstrahlung emissions were able to be neglected since the number of
electrons is sufficiently small. Since the synchrotron emission from
the secondary electrons was not able to explain the X-ray emission,
the X-ray emission might originate in the synchrotron emission from
the primary electrons, as shown in Figure~\ref{fig:hadron}, or other
emission mechanisms. Since the synchrotron emission from the secondary
electrons dominates below the infrared band, the detection of the
emission in the radio to infrared bands could support the hypothesis
that the VHE gamma-ray emission is produced by the neutral-pion
decay. Because the {\it Spitzer} upper limit was above the predicted
flux of synchrotron emission, more detailed observations are needed.


%
%
To discuss the possible association between the SNR and Pismis~22, we
estimated the SNR age using the distance to Pismis~22 with the
assumption that a supernova explosion occurred at the position of
Pismis~22 and the shock front has now reached at the VHE gamma-ray
peak position. By using the same equation (2) in \citet{yamazaki_old},
the age was obtained to be $3\times 10^{2}$~yr for a distance of 1~kpc. 
Since the lower limit of a magnetic field was obtained to be 
$B\ge $200~$\mu$G as above, the synchrotron cooling time of 
100~TeV electrons decreased to 1~yr.
For a distance of 1~kpc, the required total energy of protons $W_p$ could 
be reduced if the number density of ambient matter $n_p$ is the same 
as for a distance of 10~kpc. 
If the total energy of protons $W_p$ is fixed to be 10$^{50}$ ergs,
the required density of ambient matter is reduced to
$n_{p}=1~p$~cm$^{-3}$ for a distance of 1~kpc.
This value was comparable with the typical number density in the 
interstellar field and does not contradict the fact that no 
obvious molecular cloud was found in the NANTEN $^{12}$CO(J=1-0)
survey data. 
The contribution of emissions from secondary electrons was shown 
in Figure~\ref{fig:hadron}, assuming the injection time of $3\times
10^{2}$~yr with a magnetic field of 200~$\mu$G. 
The synchrotron emission from
the secondary electrons was not able to explain the X-ray spectrum of
Suzaku source A. The X-ray emission might originate in the synchrotron 
emission from the primary electrons, or other emission mechanisms. 
Since the flux of synchrotron emission from the secondary
electrons in the radio to infrared bands was 
lower than that of the case of a distance of 10~kpc, as shown in
Figure~\ref{fig:hadron}, a determination of the spectrum below the infrared band 
is a key to reveal the origin of the SNR. Additionally, 
an observation of thin thermal plasma
in the X-ray band will provide useful information such as plasma
temperature or chemical abundances. For example, the detection of high
abundance of $\alpha$-elements (O, Mg, Si, S, Ca, and Ti) compared to
that of iron, which is expected in a massive star explosion
\citep{kobayashi}, may support the SNR scenario. 
In fact, a recent {\it Suzaku} observation of the open cluster
Westerlund~2 detected metal-rich thermal emission, suggesting that the
diffuse X-ray and VHE gamma-ray emission may have originated from a
hypernova remnant \citep{fujita}.

\subsection{Stellar Wind Scenario}
%
%
The VHE gamma-ray emission might have been produced by hadrons
accelerated in winds from massive stars in
Pismis~22~\citep{Voelk_1982,Bednarek}. A fraction of the stellar wind energy can
be transferred to relativistic particles. Assuming that the shock
acceleration generates a single power-law spectrum of primary
particles, we can apply the discussion made in the SNR scenario. We
discuss the energetics for the hadronic origin here to produce the
observed gamma-ray emission.  A single O-type star loses mass at a
rate of $\dot{M}=10^{-6}M_{\odot}$ per year with a stellar wind
velocity of $\sim1500$~km~s$^{-1}$ \citep{castor}. The rate of kinetic
energy emitted from the single star is $7\times10^{35}$ erg
s$^{-1}$. If we assume an energy conversion efficiency to particle
acceleration of 5~\%, which is the maximum efficiency adopted for a
hadronic model in \citet{Bednarek}, an ambient matter density of
$100~p$~cm$^{-3}$, a distance of 1\,kpc, and an age of 40\,Myr, two
O-type stars are required in Pismis~22 to produce the observed VHE
gamma-ray spectrum in the hadronic scenario. 
However, no obvious molecular cloud has been found at this distance in the NANTEN data, as described above.

%
%

\subsection{PWN Scenario}
%
%
PWNe are the largest class of identified Galactic VHE gamma-ray
sources. We discuss the possibility of HESS~J1614$-$518 being a PWN in
this subsection.

As calculated in \citet{rowell}, the spin-down luminosity of each of the five 
nearby pulsars (Fig. \ref{fig:map3}) is smaller than the TeV gamma-ray
luminosity. Thus, none of already known pulsars can be associated with
HESS~J1614$-$518. There is also the possibility that an undiscovered
pulsar with a high spin-down power, sufficient to explain the observed
gamma-ray luminosity, might be located in the vicinity of Suzaku
source~A or source~B. We estimated the pulsar age to be 24~kyr and 23~kyr for Suzaku source A and B, respectively,  using the correlation
between pulsar age and the ratio of gamma-ray flux and X-ray flux
from the PWN \citep{Mattana}. For this relatively old age, we should
apply the time-evolving electron injection model (section~1). We
applied this model to HESS~J1614$-$518, following the calculation in
\citet{Higashi_2008} applied to HESS~J1804-216. Although this model
can explain the large ratio between the X-ray and TeV gamma-ray fluxes
assuming a single power-law electron distribution, the model showed
a very large discrepancy with the sub-GeV flux observed with {\it
Fermi}, the model curve is 30 times larger than the observed flux at 0.1~GeV. Thus, the time-evolving electron injection model with a
single power-law electron distribution was rejected. The MeV/GeV component could arise from the different mechanism than the TeV emission.

Although the present sensitivity in the radio band may not be 
sufficient to detect this unknown pulsar, further observations in the GeV
band could detect a radio quiet pulsar like Geminga, which was
detected with {\it CGRO} EGRET \citep{egret}, or sixteen previously
unknown pulsars which were recently discovered with {\it Fermi}
\citep{fermi_pulsar}. In addition, future X-ray observations could
detect pulsed emission from Suzaku source~A or source~B. 
Given these detections, emission models for PWNs
\citep[e.g.,][]{Nakamori_2, Tanaka, Slane,2010MNRAS.tmp.1423B}, using 
a broken power-law distribution of electrons,
might be able to reveal that the VHE gamma-ray emission of this
unidentified source originates from a PWN.

\section{Conclusion}

The observation of HESS J1614$-$518 with the CANGAROO-III telescopes
confirms the VHE gamma-ray emission reported by H.E.S.S. The
differential energy spectrum can be fitted with a single power
law:$(8.2\pm2.2_{stat}\pm2.5_{sys})\times10^{-12}\times(E/1
$TeV$)^{-\gamma}$ cm$^{-2}$s$^{-1}$TeV$^{-1}$ with a photon index
$\gamma$ of $2.4\pm0.3_{stat}\pm0.2_{sys}$. We discuss the possible
counterparts for this object using the results of 
observations with {\it Suzaku} and {\it Fermi}. For the SNR scenario, a one-zone
leptonic model was not able to account for the observed SED. 
Hadronic models gave a good reproduction of the SED and the typical SNR
explosion energy of $\sim10^{51}$\,ergs is able to supply the total
energy of protons. 
Since the required number densities of the ambient matter were 
$n_{p}=1~p$~cm$^{-3}$ and $n_{p}=100~p$~cm$^{-3}$ for a distance 
to the SNR of 1~kpc and 10~kpc, respectively, detailed molecular 
observations could determine whether the SNR originated from 
Pismis~22 (d$\sim$1~kpc) or a farther distance.
As there were also differences in the spectrum of the emission 
from the secondary electrons, a determination of the spectrum below the 
infrared band would help determine the likelihood of an SNR origin.
For the PWN scenario, the nearby
known pulsars are not responsible 
since the spin-down powers are insufficient to produce the
observed TeV gamma-ray luminosity. Further observations to search 
a pulsar are necessary to investigate the PWN scenario. 
For the stellar wind scenario,
Pismis~22 was required to contain two O-type stars through
its entire age from energetics considerations. However, the required 
number density of the ambient matter of $n_{p}=100~p$~cm$^{-3}$ may 
not be consistent with the results of the NANTEN observations.

To identify HESS~J1614$-$518, more detailed multiwavelength
observations are required. To discuss the stellar wind origin in more
detail, a determination of the number of OB stars is necessary. For
the SNR scenario, the ultra-high energy resolution of the SXS onboard
{\it Astro-H} could detect line emissions with a high abundance of
$\alpha$-elements compared to that of iron, which would indicate that
HESS~J1614$-$518 is an SNR. To show that the Suzaku source~B is an
AXP, supporting the SNR or PWN scenario, a high time-resolution 
X-ray observation is needed to detect a pulsed signal from the source.
More detailed gamma-ray spectroscopy with {\it Fermi}, and the
Cherenkov Telescope Array \citep{CTA} could determine the origin of
the accelerated particles.


\acknowledgments

The authors would like to thank H.~Matsumoto for providing useful
information about the {\it Suzaku} data, H. Yamamoto for 
fruitful discussion about the CO data, and S.~Nishiyama and
T.~Sawano for helping us use the {\it Spitzer} data. 
We also thank the anonymous referee for helpful comments.
This work was supported by a Grant-in-Aid for Scientific 
Research by the Japan Ministry of
Education, Culture, Sports, Science and Technology (MEXT), the
Australian Research Council, and the Inter-University Research Program
of the Institute for Cosmic Ray Research. The work is also supported
by Grant-in-Aid for the Global COE Program ``The Next Generation of
Physics, Spun from Universality and Emergence" from MEXT of Japan.  We
thank the Defence Support Center Woomera and BAE systems and
acknowledge all the developers and collaborators on the GALPROP
project.  T.\ Mizukami was supported by Japan Society for the
Promotion of Science Research Fellowships for Young Scientists.

\end{document}